\documentclass[aps,prl,twocolumn,ams,amsmath,showpacs,preprintnumbers,superscriptaddress]{revtex4}
\usepackage{tabularx}
\usepackage{bm}
\usepackage{subfigure}
\usepackage{euscript}
\usepackage{epsfig,psfrag,subfigure}
\usepackage{graphicx,psfrag,subfigure}
\usepackage{color}
\usepackage{amsmath}
\usepackage{amsfonts}
\usepackage{exscale}
\usepackage{amsbsy}
\usepackage{subfigure}

\renewcommand{\r}{\mathbf r}

\def\be{\begin{equation}}
\def\ee{\end{equation}}
\def\ba{\begin{eqnarray}}
\def\ea{\end{eqnarray}}

\begin{document}

\title{Klein tunneling and magnetoresistance of \textit{p-n} junctions in Weyl semimetals}	
	
	\author {Songci Li}
	
	\affiliation{Department of Physics, University of Washington, Seattle, Washington 98195, USA}
	
	\author{A. V. Andreev}
	
	\affiliation{Department of Physics, University of Washington, Seattle, Washington 98195, USA}
	
	\author {B. Z. Spivak}
	
	\affiliation{Department of Physics, University of Washington, Seattle, Washington 98195, USA}

\date{\today}

\begin{abstract}
We study the zero temperature conductance and magnetoconductance of ballistic \textit{p-n} junctions in Weyl semimetals. Electron transport is mediated by Klein tunneling between \textit{n}- and \textit{p}- regions. The chiral anomaly that is realized in Weyl semimetals plays a crucial role in the magnetoconductance of the junction. With the exception of field orientations  where the angle between $\mathbf{B}$ and the junction plane is small, magnetoconductance is positive and linear in $B$ at both weak and strong magnetic fields.  In contrast, magnetoconductance in conventional \textit{p-n} junctions is always negative. 
\end{abstract}


\maketitle

A theory of low temperature resistance and magnetoresistance  (MR) of \textit{p-n} junctions in conventional semiconductors was developed long ago~\cite{keldysh,Haering,Aronov}. The junction conductance is determined by tunneling processes of electrons between the conduction and valence bands in the presence of the built-in electric field of the junction. In this case the MR is positive, and becomes exponentially large at strong magnetic fields $B$. Two- and one-dimensional \textit{p-n} junctions in semiconductors with a gapless Dirac spectrum $\varepsilon_{\mathbf{p}} =\pm v|\mathbf{p}|$ ($v$ is the velocity of electrons) can be realized in graphene~\cite{Cheianov,Levitov,Fogler}, armchair carbon nanotubes~\cite{Anton,WeiChen} and on the surface of topological insulators~\cite{ShouchengZhang}. The main difference with conventional semiconductors is that in the gapless case the junction conductance is dominated by Klein tunneling; electrons near normal incidence are transmitted through the junction without backscattering. As a result, at $B=0$ the conductance of a graphene \textit{p-n} junction is proportional to the square root of the built-in electric field $E$ of the junction,
$
G\sim \frac{e^2}{h}\frac{W}{l_E}
$. Here $W$ is the width of the junction  and  $l_E=\sqrt{\hbar v/|e|E} $ is the characteristic length determined by the built-in electric field $E$.
In a finite magnetic field perpendicular to the graphene sheet the MR of the junction is positive~\cite{Levitov} and becomes exponentially large at large $B$.

Recently a new class of three-dimensional materials (Weyl semimetals) was discovered~\cite{TurnerVishwanath,BurkovBalents,HalaszBalents,BurkovHookBalents,VafekVishwanath,Qiang Li_Kharzeev,Huang,Xu1,Xu2,JunXiong_Ong,Jun Xiong,Hasan,Xiaojun Yang,Cai-ZhenLi,ChenZangEnzeZhangYanwenLiu,Drezden}, in which dynamics of  low energy electrons in valley $i$ may be described by a gapless Dirac Hamiltonian
\begin{equation}\label{eq:Dirac}
 H_i= \chi_i v\,\boldsymbol{\sigma}\cdot \mathbf{p} +\Delta_i + U(\r).
\end{equation}
Here $\chi_i =\pm 1$ is the valley chirality, $\boldsymbol{\sigma}=(\sigma_x,\sigma_y,\sigma_z) $ are the Pauli matrices, $\mathbf{p}$ is the momentum measured from the Weyl node, $\Delta_i$ is the energy offset of the Weyl node from the chemical potential in an undoped crystal,  and $U(\r)$ is the potential energy. Due to the Nielsen-Ninomiya theorem~\cite{NielsonNyninomiya} the number of the Weyl nodes, $g$, in the Brillouin zone must be even, and the number of opposite chirality nodes should be equal. The stability of the Weyl nodes is related to the fact that the flux of Berry curvature through a closed surface  surrounding the node is quantized.
Since the time reversal symmetry requires  the Berry curvature to be an odd function of momentum and inversion symmetry requires it to be even, Weyl nodes can only exist in crystals with either broken inversion or time reversal symmetry. In the former case the minimal number of Weyl nodes is four, while in the latter case it is two. An interesting property of Weyl fermions  is the existence of chiral (zeroth) Landau levels in a magnetic field. This feature is related to the chiral anomaly~\cite{Adler,Jackiw,NielsonNyninomiya} and leads to a strong anisotropic MR~\cite{NielsonNyninomiya,SonSpivak,SpivakAnton} in these materials. In this Letter we study the conductance and magnetoconductance of a \textit{p-n} junction in a Weyl semimetal. We show that the interplay between the chiral anomaly and and Klein tunneling results in negative MR of the junction.

\begin{figure}
		\begin{center}
		\includegraphics[height=2.1in,width=3.2in]{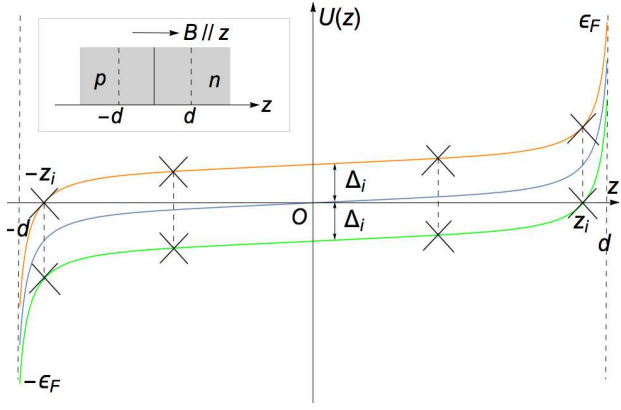}
		\caption{\label{fig:electricpotential} The sketch of the built-in electric potential (blue line) of the junction $U(z)$. The crossing points $z_i$ correspond to locations where the electron-like (green line) and hole-like (orange line) Weyl nodes cross the Fermi level.}
	\end{center}
\end{figure}

The specific geometry of the junction is shown in the inset of Fig.~\ref{fig:electricpotential}. Doping in the \textit{p}- and \textit{n}- regions creates a built-in electrostatic potential $U(z)$. Similar to graphene \textit{p-n} junctions~\cite{Cheianov} the probability of Klein tunneling in valley $i$ is determined by the value of the built-in electric field $E_{i}$ at the crossing points, $z_i$, where the electrochemical potential crosses the Weyl node, i.e. $U(z_i) + \Delta_i=0$, see Fig.~\ref{fig:electricpotential}.  Therefore we start by expressing the conductance in terms of the electric fields $E_i$ at the crossing points and then evaluate these fields by solving the corresponding nonlinear screening problem.

\textit{Conductance at zero magnetic field.} Let us consider transmission  of an electron at the Fermi level across the junction. For an electron in valley $i$ with momentum parallel to the junction plane, $\hbar\mathbf{k}_\parallel=\hbar(k_x,k_y)$, the transmission coefficient may be determined by solving an one-dimensional Dirac equation,
\begin{equation}\label{eq:Dirac_1D}
    \left(
      \begin{array}{cc}
        - i \hbar v\, \partial_z + U(z) +\Delta_i & v \hbar k_\parallel \\
        v\hbar  k^*_\parallel & i \hbar v\, \partial_z + U(z) + \Delta_i\\
      \end{array}
    \right)
    \left(
      \begin{array}{c}
        u \\
        v \\
      \end{array}
    \right)=0.
\end{equation}
Here the complex wavenumber  $k_\parallel=k_x-ik_y$ parameterizes the conserved  momentum parallel to the junction plane. We will assume that the dimensionless coupling constant $\alpha =g e^2/{\hbar v}$ is small. In this case, in the region relevant  for Klein tunneling, which is of order $l_{E_i}=\sqrt{\hbar v/|e|E_i}$ near the crossing points, the potential can be approximated by $U(z)+ \Delta_i=-e E_i (z-z_i)$. In such case the transmission coefficient is well known
\begin{equation}\label{eq:T_p}
    \mathcal{T}_i(|k_\parallel|)=  \exp\left(-\pi |k_\parallel|^2 l^2_{E_i}\right).
\end{equation}
This result may be understood from a semiclassical consideration. For a given $k_\parallel$ the value of the $z$-component of the electron momentum is dictated by energy conservation, $v p_z (z)=\pm\sqrt{\left[eE_i(z-z_i)\right]^2 -(\hbar v |k_\parallel|)^2}$, which yields the stopping points $z_i\pm |k_\parallel|l_{E_i}$. The transmission coefficient is determined by the imaginary part of the action of the tunneling trajectory accumulated in the classically forbidden region between the stopping points, $\mathcal{T}_i(|k_\parallel|)=\exp \left(-2\,\mathrm{Im}\int p_z(z) dz/\hbar \right)=\exp\left(-\pi |k_\parallel|^2 l^2_{E_i}\right)$. This coincides with the exact result, Eq.~(\ref{eq:T_p}), according to which only electrons with small parallel momenta, $|k _\parallel|\lesssim l^{-1}_{E_i}$, have an appreciable transmission probability.

Substituting Eq.~(\ref{eq:T_p}) into the Landauer formula and summing over valleys and $k_\parallel$, we obtain the conductance of the junction
\begin{equation}
  G= \frac{e^2}{ h}\sum_i\frac{S}{ (2\pi l_{E_i})^2}, \label{eq:conductancezerofield}
\end{equation}
where $S$ is the area of the junction.

\begin{figure}
		\begin{center}
		\includegraphics[height=1.9in,width=3.2in]{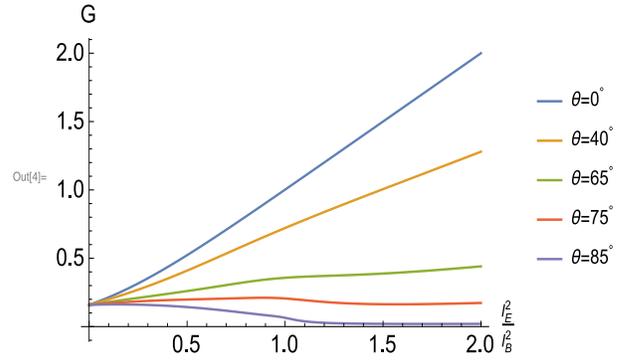}
		\caption{\label{fig:magnetozerotemp} The magnetic field dependence of the conductance $G$ at different  angles $\theta$ between the magnetic field and the normal to the junction plane.  $G$ is measured in the units of the $\frac{e^2}{h}\frac{S}{2\pi l^2_E}$.}
	\end{center}
\end{figure}

\emph{Magnetoconductance.} Next we consider the magnetic field dependence of the junction conductance $G(B)$ at zero temperature for a magnetic field perpendicular to the plane of the junction. In the vicinity of the crossing points the electron Hamiltonian has the form $H_i = v\,\boldsymbol{\sigma}\cdot\left(-i\hbar\boldsymbol{\nabla}-\frac{e}{c}\mathbf{A}\right)-e E_i z$. Using the Landau gauge for the vector potential, $\mathbf{A}=(0, Bx, 0)$,
we look for the energy eigenstates in the form $\psi^T=e^{i k_y y} (u(x,z),v(x,z))$. The spinor amplitudes $u$ and $v$ satisfy
the Dirac equation
\begin{equation}\label{eq:hamiltonian}
	\hbar v \left(
	\begin{array}{cc}
		\frac{\partial}{\partial z}-i\frac{z}{l^2_{E_i}}& \frac{\partial}{\partial x} - \frac{x-x_0}{l_B^2} \\
		\frac{\partial}{\partial x} +\frac{x-x_0}{l_B^2} & -\frac{\partial}{\partial z}-i\frac{z}{l^2_{E_i}} \\
	\end{array}
	\right)
\left(
  \begin{array}{c}
    u \\
    v \\
  \end{array}
\right)=0,
\end{equation}
with $l_B=\sqrt{\hbar c/|e|B}$ being the magnetic length and $x_0=k_yl^2_B$. The solutions have the form $(u,v)=\left(\phi_{n-1}(x) \tilde u_{n-1}(z),\phi_n(x) \tilde v_n(z)\right)$, where $\phi_n(x)$ are wavefunctions of the $n$-th Landau level. The amplitudes $\tilde{u}$ and $\tilde{v}$ obey the differential equation
\begin{equation}
	\hbar v
	\left(
	\begin{array}{cc}
		\frac{\partial}{\partial z}-i\frac{z}{l^2_{E_i}}& \frac{\sqrt{2n}}{l_{B}} \\
		\frac{\sqrt{2n}}{l_{B}} & -\frac{\partial}{\partial z}-i\frac{z}{l^2_{E_i}} \\
	\end{array}
	\right)
	\left(
	\begin{array}{c}
	\tilde{u}_{n-1}(z)\\
	\tilde{v}_n(z)
	\end{array}
	\right)=0
	. \label{eq:tildehamiltonian}
\end{equation}
 Note that in addition to ``conventional" Landau levels there is one chiral, $n=0$, Landau level (in this case $\tilde{u}_{n-1}=0$).
Since Eq.~(\ref{eq:tildehamiltonian}) coincides with Eq.~(\ref{eq:Dirac_1D}) for a quantized value of the parallel momentum,
 $|k_{\parallel, n}|=\sqrt{2n}/l_B$, the transmission coefficient for the $n$-th Landau level may be obtained by substituting $| k_{\parallel, n}| = \sqrt{2n}/l_B$ in Eq.~(\ref{eq:T_p}).
 \begin{equation}\label{eq:T_B}
\mathcal{T}_{n,i}=\exp\left(-2 \pi n \, \frac{l ^{2}_{E_i}}{l^2_B}\right).
\end{equation}
Substituting Eq.~(\ref{eq:T_B}) into the Landauer formula, summing over the Landau levels  and accounting for their degeneracy, $S/(2\pi l_B^2)$,  we get the magnetic field dependence of the conductance
\begin{eqnarray}
G(B) &=& \frac{e^2}{h} \frac{S}{2\pi l^2_B}\sum_{i}\frac{1}{1-e^{-2\pi l^{2}_{E_i}/l^2_B}}\,, \label{eq:conductancenonzerofield}
\end{eqnarray}
which is plotted in Fig.~\ref{fig:magnetozerotemp} ($\theta=0$ curve). As expected, at  $B\to 0$ the above expression reproduces the zero field result,  Eq.~(\ref{eq:conductancezerofield}).

It follows from Eq.~(\ref{eq:conductancenonzerofield}) that  $G(B)$ is a monotonically increasing function of the magnetic field. Note that the magnetoconductance is a linear  function of the magnetic field, $\delta G(B)=G(B)-G(0) \sim B$,  at both weak $(l_B \gg l_E)$ and strong $(l_E \gg l_B)$ fields. The positive magnetoconductance (or negative MR) is a signature of the chiral anomaly in Weyl materials~\cite{NielsonNyninomiya,SonSpivak,SpivakAnton}. At strong fields, $l_B\ll l_E$, the conductance is determined entirely by the electrons in the chiral ($n=0$) Landau levels, which move between the \textit{p}- and \textit{n}- regions without backscattering. In this case the positive magnetoconductance is due to the linear in $B$ growth of Landau level degeneracy.

The results (\ref{eq:conductancezerofield}) and (\ref{eq:conductancenonzerofield}) assume absence of scattering and inter-valley electric or magnetic breakdown. They hold provided the electron mean free path exceeds $l_E$ and $l_B$, and the magnetic field and the built-in electric field are not too strong; $\Delta K \gg l_{E}^{-1}, l_B^{-1}$ (here  $\Delta K$ is the momentum difference between the Weyl nodes). The electric fields $E_i$  must be determined by solving a nonlinear screening problem inside the junction.

\textit{ Tilted magnetic field.} In the general situation, in which the magnetic field makes an angle $\theta$ with the normal to the junction plane, the electron transmission problem can be solved analytically. The resulting conductance of the junction is obtained in the appendix~\cite{Supplemental} and is given by,
\begin{equation}\label{eq:conductancetiltedfield}
	G(B)=\frac{e^2}{h}\frac{S\cos\theta}{2\pi l_B^2}\sum_i\frac{\cos\alpha_i(\theta)}{1-\exp(-2\pi\sqrt{\frac{|\lambda_{i-}|}{\lambda_{i+}}})},
\end{equation}
where
\begin{eqnarray*}
&&\lambda_{i\pm}=\frac{1}{2}\sqrt{\left(\frac{l^4_B}{l^4_{Ei}}-1\right)^2+\frac{4\cos^2\theta\,l^4_B}{l^4_{Ei}}}\pm\frac{1}{2}\left(\frac{l^4_B}{l^4_{Ei}}-1\right), \\
&& \tan 2\alpha_i(\theta)=\frac{\sin2\theta}{\cos2\theta+l^4_B/l^4_{Ei}}.
\end{eqnarray*}
For $\theta =0$ this expression reproduces Eq.~(\ref{eq:conductancenonzerofield}). 
Magnetoconductance remains positive and linear in $B$ at both low and high fields for most tilting angles $\theta$, see Fig.~\ref{fig:magnetozerotemp}. For $\theta\gtrsim 70^{\circ}$ magnetoconductance becomes non-monotonic and develops a shoulder-like  feature at $l_B/l_E \sim 1$. The latter arises because at $l_B/l_E \gg 1$ Klein tunneling occurs  along the $z$-axis, whereas at $l_B/l_E \ll1$  tunneling occurs in the direction of the magnetic field. As a result, for $\theta$ close to $90^{\circ}$  the apparent area of the junction, available for tunneling sharply decreases as the increasing magnetic field passes $l_B=l_E$ .

\textit{Evaluation of the built-in electric field.} For simplicity, below we assume that the offsets in the electron-like and hole-like valleys are equal in magnitude, $\Delta_i=\pm \Delta$. The corresponding density of states  has the form $\nu(\varepsilon)=g(\varepsilon^2+\Delta^2)/\pi^2\hbar^3v^3$, and the number density of electrons depends on the electrostatic potential as $n(U) =-g\left(U^3 + 3\Delta^2U\right)/(3\pi^2\hbar^3v^3)$. The electrostatic potential $U(z)$ obeys the following Poisson equation,
\begin{equation}\label{eq:Poisson}
  \frac{d^2U(z)}{dz^2}=4\pi e^2 \left[-n_\mathrm{d}(z) + g \frac{U^3 + 3\Delta^2U}{3\pi^2\hbar^3v^3}\right],
\end{equation}
where $n_\mathrm{d}(z)$ is the dopant density, which we model as $n_\mathrm{d}(z)=n_0\,\mathrm{sgn}(z) \Theta(|z|-d)$ with $\Theta(x)$ being the step function.

Before presenting an analytic solution of Eq.~(\ref{eq:Poisson}) let us begin with a qualitative discussion of its essential features. Deep inside the doping regions, $|z| \gg d$, the electrostatic potential approaches constant values $ \pm \varepsilon_F $ determined by the dopant density $n_0$. In the middle of the junction $|U(z)| \ll \Delta$, and the screening is linear, with the intrinsic screening length $\kappa^{-1}=\sqrt{\pi/4\alpha}\,\hbar v/\Delta$. At $ |U(z)| \gtrsim \Delta $ screening becomes nonlinear. Since the creation of the \textit{p-n} junction requires $|U (z)| > \Delta $ one should distinguish between the following two cases: \textit{i}) moderate doping, $ \varepsilon_F \gtrsim \Delta $, and \textit{ii}) strong doping,  $\varepsilon_F \gg \Delta $. In either case we assume that the junction width $d$ exceeds the screening length in the doping region, $d \gg (\sqrt{\alpha}\,\varepsilon_F/\hbar v)^{-1}$. The magnitude $E_*$ of the electric field at the crossing points in these regimes may be estimated as follows.

\textit{i}) For moderate doping, $\varepsilon_F \gtrsim \Delta$, the crossing points are located within the screening length $\kappa^{-1}$ from the boundary of the doping regions, and the electric field at the crossing points may be estimated as $E_*\sim \varepsilon_F \kappa/|e|$. Here  we assume that Fermi energies in different valleys are of the same order $\varepsilon_{F}$. Using Eq.~(\ref{eq:conductancezerofield}) the conductance can be estimated as
\begin{equation}\label{eq:G_0_estimate_linear}
    G(0)\approx\frac{e^2}{2\pi h} \frac{g S}{2\pi}k_F\kappa, \quad k_F=\frac{\varepsilon_F}{\hbar v}.
\end{equation}
Note that the conductance turns out to be independent of the junction width $d$.

\textit{ii}) For strong doping, $\varepsilon_F \gg \Delta$,  near the boundary with the doping region, $d -| z| \ll \kappa^{-1}, d$, the last term in Eq.~(\ref{eq:Poisson}) may be neglected and the solution (on the doping-free side) acquires a simple form,
\[
U(z)\approx A/( d+z_0- |z |).
\]
Since inside the doping region $|U(z)|\sim \varepsilon_F$ and the screening length is $\sim (k_F \sqrt{\alpha})^{-1}$  continuity of the potential and electric field at $|z|=d$ yields  $|A| \sim v/\sqrt{\alpha}$, and $z_0 \sim 1/(\sqrt{\alpha} k_F)$.  Thus the locations of the crossing points, $|U(z_*)|=\Delta$,  may be estimated as $d-|z_*| \sim \mathrm{min}\{\kappa^{-1},d\}$, and the electric field in them as, $E_*\sim \hbar v/ |e|\sqrt{\alpha}\,\mathrm{min}\{\kappa^{-2}, d^2\}$.  This results the following estimate for the junction conductance,
\[
G\sim \frac{e^2}{2\pi h}  \frac{ g S}{\sqrt{\alpha}\,\mathrm{min}\{\kappa^{-2},d^2\}}.
\]
Note that at strong doping the conductance becomes independent of the doping level $\varepsilon_F$.

Let us now turn to the quantitative treatment of the nonlinear screening problem. The first integral of the Poisson equation (\ref{eq:Poisson}) can be cast in the following dimensionless form,
\begin{subequations}
\label{eq:first_integral_U}
\begin{eqnarray}
	\!\tilde{U}^2_\zeta \!&\!=\!&\!\left(\tilde{U}-1\right)^2\left(\tilde{U}^2+2\tilde{U}+3+6\delta^2\right), \zeta>\zeta_d, \label{eq:first_integral_U_a}\\
	\!\tilde{U}^2_\zeta\!&\!=\!&\!\tilde{U}^4+6\delta^2\tilde{U}^2+\tilde{E}_c^2,\, 0<\zeta<\zeta_d, \label{eq:first_integral_U_b}
\end{eqnarray}
\end{subequations}
where $\tilde{U}=U/\varepsilon_F$, $ \delta=\Delta/\varepsilon_F$, $\zeta=\sqrt{2\alpha/3\pi}k_F z$ and $\tilde{E}_c=\sqrt{3\pi/2\alpha}\,|e|E_c/(k_F\varepsilon_F)$ are, respectively, the dimensionless electrostatic potential, energy offset, coordinate, and electric field at the center of the junction. Finally, $\tilde{U}_\zeta$ denotes the first derivative of $\tilde{U}$ with respect to $\zeta$ and $\epsilon_F$ is related to the dopant density by $n_0=4\alpha(1+3\delta^2)\,k_F\varepsilon^2_F/(3\pi^2e^2)$.

The solution of Eqs.~(\ref{eq:first_integral_U_a}) inside the doping region $\zeta>\zeta_d$ is given by
\begin{equation}
\label{eq:sol_U_a}
\tilde{U}=1-\frac{3\sqrt{2}\,(1+\delta^2)}{\sqrt{2}+\sqrt{1+3\delta^2}\sinh\sqrt{6(1+\delta^2)}(\zeta-\zeta_0)}.
\end{equation}
The solution of Eq.~(\ref{eq:first_integral_U_b}) in the doping-free region is given by
\begin{equation}
\label{eq:sol_U_b} 
\tilde{U}=-ia_-\,sn\left(ia_+\zeta,k\right),
\end{equation}
where $sn(u,k)$ is the Jacobi elliptic function~\cite{Erdelyi}, and the parameters $a_\pm$ and $k$ are given by
\begin{equation}
\label{eq:a_pm_def}
a_{\pm}=\sqrt{3 \delta^2\pm\sqrt{9\delta^4-\tilde{E}_c^2}},\quad k=\frac{a_-}{a_+}.
\end{equation}

The integration constants $\zeta_0$ and $\tilde{E}_c$ in Eqs.~(\ref{eq:sol_U_a}) and (\ref{eq:sol_U_b}) are determined from the following  equations, which express  the continuity of the potential $\tilde{U}$ and its derivative at the boundary of the doping region ($\zeta=\zeta_d$),
\begin{subequations}
	\begin{eqnarray*}
		\frac{3\sqrt{2}\,(1+\delta^2)}{\sqrt{2}+\sqrt{1+3\delta^2}\sinh\left[\sqrt{6(1+\delta^2)}(\zeta_d-\zeta_0)\right]}-1&=&ia_-s_d,\\
		\frac{6\sqrt{3}\,(1+\delta^2)\cosh\left[\sqrt{6(1+\delta^2)}(\zeta_d-\zeta_0)\right]}{\left\{\sqrt{2}+\sqrt{1+3\delta^2}\sinh\left[\sqrt{6(1+\delta^2)}(\zeta_d-\zeta_0)\right]\right\}^2}&=&\tilde{E}_c c_d d_d.
	\end{eqnarray*}
\end{subequations}
Here the abbreviations $s_d, c_d$ and $d_d$ stand for
\[
s_d\equiv sn(ia_+\zeta_d,k),\, c_d\equiv cn(ia_+\zeta_d,k),\, d_d\equiv dn(ia_+\zeta_d,k).
\]
The dimensionless electric field at the center of the junction, $\tilde{E}_c$ can be found by solving the above equations numerically. For the  dimensionless electric field $\tilde{E}_*$ at the crossing points, $\tilde{U}(\zeta_*)=\pm\delta$,  using Eq.~(\ref{eq:first_integral_U_b}) we get
\begin{equation}\label{eq:E_crossingpoint}
	\tilde{E}_*^2=\tilde{E}_c^2+7\delta^4.
\end{equation}

At strong doping determination of the potential inside the undoped region can be further simplified. In this case both $\tilde{E}_c$ and $ \delta $ are small, and  by Eq.~(\ref{eq:a_pm_def}) so are $a_\pm$. Then the matching conditions can be satisfied only if the function $sn(ia_+\zeta,k)$ in Eq.~(\ref{eq:sol_U_b}) has a pole near the boundary with the doping region, $\zeta \approx \zeta_d$. Since in real space the location of the pole is offset from $\pm d$ by a distance of order of the screening length in the doping region, then to accuracy $1/(\sqrt{\alpha} k_F d)$ we can determine $\tilde{E}_c$ from the condition that  $sn(ia_+ \zeta, k)$ in Eq.~(\ref{eq:sol_U_b}) must have a pole at $\zeta=\zeta_d$.

The Jacobi elliptic function $sn(w,k)$ has a series of simple poles at $w=w_{mn} =2mK(k)+(2n+1)iK(\sqrt{1-k^2})$ with residues $(-1)^m/k$. Here $m$, $n$ are integers, and $K(k)=\int_{0}^{\pi/2}d\phi/\sqrt{1-k^2\sin^2\phi}$, is the complete elliptic integral of the first kind. Near the poles the dimensionless potential $\tilde{U}$ in Eq.~(\ref{eq:sol_U_b}) has the  form,
$
\tilde{U}(\zeta)\approx (-1)^{m+1}/(\zeta+i\zeta_{mn}/a_+).
$
Since $\tilde{U}$ must be real for real $\zeta$ the pole at $\zeta=\zeta_d$  corresponds to $m=n=0$. This gives the condition  that determines the dimensionless electric field $\tilde{E}_c$ at the center of the junction,
\begin{equation}\label{eq:pole_k}
  \zeta_d= \frac{1}{a_+}K\left(\sqrt{1-k^2}\right).
\end{equation}
The right hand side of this condition is real for all values of $\tilde{E}_c$. For $\tilde{E}_c < 3 \delta^2$ this is obvious since in this regime
$0< k<1$ and $a_+$ is real, see Eq.~(\ref{eq:a_pm_def}).  For $\tilde{E}_c> 3\delta^2$ the location of the pole remains real although the parameters $a_\pm$ and $k$ become complex. To see this we express $\tilde{E}_c$ in terms of a parameter $\theta$ in the form
\begin{equation}
\label{eq:E_c_theta}
\tilde{E}_c = \frac{3\delta^2}{\cosh\theta}.
\end{equation}
Here $\theta$ is real and positive for $\tilde{E}_c<3\delta^2$, and becomes purely imaginary, $\theta\rightarrow i\vartheta, 0<\vartheta<\pi/2$, for $3\delta^2<\tilde{E}_c$. The parameters $a_\pm$, and $k$ in Eq.~(\ref{eq:a_pm_def}) may be expressed in terms of $\theta$ as
$
a_{\pm}=\frac{\sqrt{3}\,e^{\pm\theta/2}}{\sqrt{\cosh\theta}}\, \delta,$ and  $k=e^{-\theta}$.
Using  the identity $K(\sqrt{1-k^2})=\frac{2}{1+k} K\left( \frac{1-k}{1+k}\right)$, see formula 13.8 (12) of Ref.~\cite{Erdelyi}, we can rewrite
Eq.~(\ref{eq:pole_k}) in the form
\begin{equation}\label{eq:pole}
\delta\,\zeta_d = \frac{\sqrt{\cosh(\theta)/3}}{\cosh(\theta/2)}K\left(\tanh\left(\frac{\theta}{2}\right)\right),	
\end{equation}
that is explicitly real for all values of the electric field $\tilde{E}_c$.
The electric field $\tilde{E}_*$ at the crossing points may be obtained using Eqs.~(\ref{eq:E_crossingpoint}) and (\ref{eq:E_c_theta}).

In the limiting case of $d\gg 1/\kappa$ (strong intrinsic screening) $\tilde{E}_c \ll 3 \delta^2$ we have $\theta \gg 1$ and  Eq.~(\ref{eq:pole}) simplifies to
$
\tilde{E}_c\approx 24\delta^2 e^{-\sqrt{6}\delta\zeta_d}=24\delta^2 e^{-\kappa d}.
$
The characteristic length $l_{E_*}$ at the crossing points can be found from Eq.~(\ref{eq:E_crossingpoint})
\begin{equation}\label{eq:l_E_strong_screening}
  l^{-2}_{E_*}\approx\sqrt{\frac{7\pi}{24\alpha}}\kappa^2\approx \frac{0.96}{\sqrt{\alpha}}\kappa^2.
\end{equation}
In the opposite limit of $d\ll 1/\kappa$ (weak intrinsic screening) we have
 $\theta= i\vartheta \to  i\pi/2$, and Eq.~(\ref{eq:pole}) yields $\tilde{E}_c\approx 2K^2(-i)/\zeta^2_d=3\pi K^2(-i)/\alpha d^2$.
The characteristic length $l_{E_*}$ corresponding to the electric field at the crossing points is given by
\begin{equation}\label{eq:l_E_weak_screening}
  l^{-2}_{E_*}\approx\sqrt{\frac{6\pi}{\alpha}}K^2(-i)\frac{1}{d^2}\approx \frac{7.45}{\sqrt{\alpha}\,d^2}.
\end{equation}

The junction conductance  (\ref{eq:conductancezerofield}) in these limiting cases  is expressed in the form,
\[
	G(0)\approx\frac{e^2}{2\pi h}\frac{g\,S}{2\pi\sqrt{\alpha}}
	\begin{cases}
	0.96\kappa^2,	& d\gg \kappa^{-1},\\
	7.45 d^{-2},	& d\ll \kappa^{-1}.
	\end{cases}
\]
As expected, at strong doping it is independent of the doping level $\varepsilon_F$.

We note that  the assumption that the potential $U(z)$ changes linearly with $z$ in the interval of order $l_{E}$ near the crossing points is justified as long as the dimensionless coupling constant is small, $\alpha \ll 1$. 

It is important to note that the MR of the junction can be significant even in the interval of magnetic fields where it can be treated semiclassically in the regions of the junction. Therefore one can neglect the magnetic field dependence of the density of states in these regions. Finally we note that the value of $E_{c}$ is unaffected by the magnetic field in all cases considered above.

We would like to discuss differences between the above negative MR in \textit{p-n} junctions and recently observed negative MR of bulk Weyl semimetals. In bulk Weyl semimetals at $\varepsilon_{F} \gg \hbar v/l_{B}$ electrons can be described semiclassically. In the latter case  the magnitude of the negative MR is quadratic in $B$~\cite{SonSpivak,SpivakAnton}. It exists only in a situation where the inter-valley relaxation time is much longer than the intra-valley one and only in certain interval of angles between the external electric and magnetic fields, and only in some (usually small) interval of angles between the external electric and magnetic fields.
In contrast, the negative MR of \textit{p-n} junction is governed by the parameter $l_E/l_B$ and is independent of the relaxation times. Both at small and large magnetic fields its magnitude is linear in $B$. 

Another way to distinguish the contribution of \textit{p-n} junction to the total negative MR of the device is to study it as a function of the bias voltage $V$ on the junction: the value of $ G(V,B)$ should exhibit characteristic asymmetry with respect to a change $V\rightarrow -V$ for diodes. 

The work of S. L. and A. A. was  supported by the U.S. Department of Energy Office of Science, Basic Energy Sciences under Award No. DE-FG02-07ER46452.

\appendix*
\section{Appendix: Derivation of the magnetoconductance in a tilted magnetic field}
\setcounter{equation}{0}    
\renewcommand{\theequation}{A.\arabic{equation}}
In this appendix we derive the expression for the conductance of the $p$-$n$ junction in a tilted magnetic field, Eq.~(\ref{eq:conductancetiltedfield}) of the main text.  Let the magnetic field be in the $x$-$z$ plane at an angle $\theta$ with the $z$-axis, $\mathbf{B}= B(\sin\theta, 0, \cos\theta)$. Near the crossing point,  $U(z)\approx- eE z$, the Dirac Hamiltonian reads
\begin{eqnarray}
	H &=&v\,\boldsymbol{\sigma}\cdot\left(-i\hbar\boldsymbol{\nabla}-\frac{e}{c}\mathbf{A}\right)-e E z, \nonumber 
\end{eqnarray}
We work in the Landau gauge $\mathbf{A}= B (0, x \cos \theta - z\sin \theta, 0)$, and consider motion of an electron with a conserved wavenumber $k_y$ along the $y$-axis. Let us make a unitary transformation which amounts to rotation about the $x$-axis in pseudospin space;
\begin{equation}
	\label{eq:unitary_transform}
	\sigma_x \to \sigma_x, \quad \sigma_y \to \sigma_z, \quad \sigma_z \to - \sigma_y.
\end{equation}
Upon this transformation the Hamiltonian becomes
\begin{eqnarray}
	\frac{H}{\hbar v }
	&=&
	\left(
	\begin{array}{cc} k_y- \frac{x\cos \theta - z\sin \theta}{l_B^2} -\frac{z}{l^2_E}& - i \partial_x + \partial_z \\
		- i \partial_x -\partial_z & -k_y+ \frac{x\cos \theta - z\sin \theta}{l_B^2}-\frac{z}{l^2_E} \\
	\end{array}
	\right). \nonumber
	\label{eq:Hamiltonian_U}
\end{eqnarray}

Let us now rescale the variables. We will measure coordinates in units of the magnetic length $l_B$, and energy in units of $\hbar v/l_B$. The rescaled Dirac equation on spinor $\psi_{\mathbf{k}}=(u_{\mathbf{k}},v_{\mathbf{k}})^T$ becomes
\begin{subequations}
	\begin{eqnarray}
		(i\partial_x+\partial_z)\!u_{\mathbf{k}}\!&\!=\!&\!\left[\!x\cos\theta -k_y-z\!\left(\sin\theta+\frac{cE}{vB}\right)\!\right] \! v_{\mathbf{k}}\label{eq:Dirac_tilted_a},\\
		(i\partial_x-\partial_z) \!v_{\mathbf{k}}\!&\!=\!&\!\left[\!k_y-x\cos\theta+z\!\left(\sin\theta-\frac{cE}{vB}\right)\!\right]\! u_{\mathbf{k}}. \label{eq:Dirac_tilted_b}
	\end{eqnarray}
\end{subequations}

We now apply the operator $ i \partial_x -\partial_z $ to  (\ref{eq:Dirac_tilted_a}) and combine the result with   (\ref{eq:Dirac_tilted_b}) multiplied by $-k_y + x \cos \theta - z\left( \sin \theta +\frac{cE}{vB} \right)$. Similarly, we apply $ i \partial_x + \partial_z$ to (\ref{eq:Dirac_tilted_b}) and combine the result with (\ref{eq:Dirac_tilted_a}) multiplied by $k_y - x \cos \theta + z\left( \sin \theta -\frac{cE}{vB} \right)$. This yields the system of second order partial differential equations,
\begin{equation}
	\label{eq:h-matrix}
	\hat{h}\left(
	\begin{array}{c}
		u_{\mathbf{k}}\\
		v_{\mathbf{k}}
	\end{array}
	\right)=\left[\sin\theta\sigma_x-\left(\cos\theta-i\frac{c E}{v B}\right)\sigma_y\right]\left(
	\begin{array}{c}
		u_{\mathbf{k}}\\
		v_{\mathbf{k}}
	\end{array}
	\right),
\end{equation}
where $\hat{h}$ is given by
\begin{equation}
	\label{eq:h_def}
	\hat h = \left( \partial_x^2 + \partial_z^2 \right)  -\left[ (k_y-x \cos \theta + z\sin \theta)^2 - z^2 \left(\frac{cE}{vB}\right)^2 \right].
\end{equation}
Diagonalizing the matrix in the right hand side of Eq.~(\ref{eq:h-matrix}) and denoting its eigenvalues by 
\begin{equation}
	\label{eq:beta_pm}
	\beta_{\pm}=\pm i \sqrt{ \left(\frac{cE}{vB}\right)^2- 1 + 2 i \cos \theta \, \frac{cE}{vB}},
\end{equation}
we get two Schr\"{o}dinger equations for the appropriate linear combinations of $u_{\mathbf{k}}$ and $v_{\mathbf{k}}$,
\[
\hat h\, \varphi_\pm =\beta_{\pm}\varphi_\pm.
\]

Changing the coordinates from $x,z$ to $\xi,\zeta$ via
\begin{equation}
	\left(
	\begin{array}{c}
		x-\frac{k_y}{\cos\theta}\\
		z
	\end{array}
	\right)=
	\left(
	\begin{array}{cc}
		\cos\alpha & \sin\alpha\\
		-\sin\alpha & \cos\alpha
	\end{array}
	\right)
	\left(
	\begin{array}{c}
		\xi\\
		\zeta
	\end{array}
	\right),
\end{equation}
where the rotation angle $\alpha$ is given by 
\begin{equation}
	\label{eq:alpha}
	\tan 2\alpha=\frac{\sin 2\theta}{\cos 2\theta+\left(\frac{cE}{vB}\right)^2} ,
\end{equation}
we can write the Schr\"{o}dinger operator $\hat h$ in the form 
\[\hat{h}=\partial^2_\xi - |\lambda_-|\xi^2+\partial^2_\zeta+\lambda_+\zeta^2,
\] 
where
\begin{equation}
	\label{eq:lambda_pm}
	\lambda_{\pm}=\frac{1}{2}\left[\left(\frac{cE}{vB}\right)^2- 1 \pm\sqrt{\left[\left(\frac{cE}{vB}\right)^2-1\right]^2+4\cos^2\theta  \left(\frac{cE}{vB}\right)^2}\right].
\end{equation}
Rescaling the coordinates $\xi$ and $\zeta$ to $\tilde\xi=|\lambda_-|^{1/4}\xi$ and $\tilde\zeta=\lambda^{1/4}_+\zeta$ we we reduce the Shr\"{o}dinger operator $\hat h$ to the simple form $\hat h=\sqrt{|\lambda_-|}\left(\partial^2_{\tilde\xi}-{\tilde\xi}^2\right)+
\sqrt{\lambda_+}\left(\partial^2_{\tilde\zeta}+{\tilde\zeta}^2\right)$.

The solutions corresponding to the scattering problem may be written in the form
$\varphi(\tilde\xi,\tilde\zeta)=f(\tilde\xi)g(\tilde\zeta)$, where $f(\tilde\xi)$ 
has the form
$
f(\tilde\xi)=e^{-{\tilde\xi}^2/2}H_n(\tilde\xi), \quad n=0,1,2\cdots
$.
Here $n$ is the Landau level index and $H_n(x)$ is the Hermite polynomial. The function  $g(\tilde\zeta)$  satisfies  the Weber equation 
\begin{equation}
	\label{eq:Weber}
	\frac{d^2 g}{d{\tilde\zeta}^2}+\left[{\tilde\zeta}^2-(i+q^2_n)\right]g(\tilde\zeta)=0,
\end{equation}
with 
\[
q^2_n=2n\sqrt{\frac{|\lambda_-|}{\lambda_+}}-\frac{\beta_{\pm}\mp\sqrt{|\lambda_-|}}{\sqrt{\lambda_+}}-i.
\]
Eq.~(\ref{eq:Weber}) describes scattering of a nonrelativistic particle at an inverted parabolic potential. The corresponding transmission amplitude is given by $\exp(-\pi q^2_n/2)$.
Using the fact that $\mathrm{Re} \beta_\pm =\pm\sqrt{\lambda_-}$
we get the transmission coefficient in the form
\begin{equation}
	\label{eq:T_n}
	\mathcal{T}_n=\exp\left(-2\pi n \, \sqrt{\frac{|\lambda_-|}{\lambda_+}} \right).
\end{equation}
Accounting for the Landau Level degeneracy and the number of Weyl nodes, we get the junction conductance in the form of Eq.~(9) in the main text,
\begin{eqnarray}
	\label{eq:conductance_tilted}
	&&G=\frac{e^2}{h} \sum_i\frac{S\cos\theta}{2\pi l_B^2}\, \frac{\cos\alpha_i}{1-e^{-2\pi \sqrt{|\lambda_{i-}|/\lambda_{i+}}}},  \\
	&&\tan2\alpha_i=\frac{\sin2\theta}{\cos2\theta+(cE_i/vB)^2}.
\end{eqnarray}
Here $i$ labels the quantities pertaining to the crossing point for the $i$-th Weyl node. 
The presence of  $\cos\alpha$  accounts for the fact that  Klein tunneling occurs along $\zeta$-axis which makes the angle $\alpha$ with the $z$-axis by $\alpha$. For a magnetic field normal to the junction ($\theta=0$) this equation reproduces Eq.~(\ref{eq:conductancenonzerofield}) in the main text.

\end{document}